\documentclass[11pt,twoside]{article}

\usepackage{asp2006}
\usepackage{epsf}
\usepackage{psfig}
\usepackage{lscape}

\begin{document}
\title{Disks, young stars, and radio waves: the quest for forming
planetary systems}
\author{Claire J. Chandler and Debra S. Shepherd}
\affil{National Radio Astronomy Observatory, PO Box O, Socorro, NM
87801, USA}

\begin{abstract}

Kant and Laplace suggested the Solar System formed from a rotating gaseous
disk in the 18th century, but convincing evidence that young stars are
indeed surrounded by such disks was not presented for another 200 years.
As we move into the 21st century the emphasis is now on disk formation,
the role of disks in star formation, and on how planets form in those
disks.  Radio wavelengths play a key role in these studies, currently
providing some of the highest spatial resolution images of disks, along
with evidence of the growth of dust grains into planetesimals.  The future
capabilities of EVLA and ALMA provide extremely exciting prospects
for resolving disk structure and kinematics, studying disk chemistry,
directly detecting proto-planets, and imaging disks in formation.

\end{abstract}

\section{Introduction}

The search for forming planetary systems is ultimately a search for our
origins.  How do the gas and dust in molecular clouds evolve into rocky
and gas giant planets, and how common are planetary systems like the
Solar System?  Radial velocity searches for extra solar planets indicate
that the Solar System might actually be unusual, but such searches
are only just achieving the sensitivities needed to detect the Solar
System-like planetary systems.  For several reasons it is much easier
to study the disks from which planetary systems are expected to form,
and this may shed light on the environment and processes taking place
in the early Solar System.

The formation of disks at some stage during the star formation process
is made inevitable by one of several ``angular momentum problems''
in star formation.  Specific angular momentum is imparted to molecular
clouds by differential Galactic rotation, and many orders of magnitude
must be lost before a planetary system can form.  The outcome of the
star formation process for solar-type stars is therefore a T Tauri star,
descending the Hayashi track towards the main sequence in the HR diagram,
surrounded by a disk.  Massive stars reach the main sequence while still
deeply embedded and accreting.

A combination of radiative heating by the central star, accretion, and
heating by the interstellar radiation field results in temperatures for
disks on Solar System size scales of 10 to 20~K for Solar-type stars.
The bulk of the disk mass is therefore best traced by millimeter and
submillimeter wavelength emission.  Indeed, the first convincing
evidence for the ubiquity of disks surrounding T Tauri stars
arose from a single-dish survey for 1.3~mm continuum emission by
\citet{chandler:Beckwith1990}.  Although those measurements were not able
to resolve the emission the flux densities detected, and corresponding
dust column densities, implied that the dust had to be distributed in
flattened structures in order to explain the low optical extinctions to
the central stars.  The rest of this review focuses on the role of radio
interferometry in spatially resolving the emission from circumstellar
disks, and in understanding the disk properties.

\section{Dust emission from disks}

Dust emission at centimeter through to submillimeter wavelengths has
played a vital role in establishing the presence of circumstellar disks
and for providing evidence of grain growth.  The radiative transfer of
the dust emission is relatively simple compared with that of molecular
lines (with caveats regarding the chemical composition and shape of
the dust grains), and may be the only tracer of mass in regions where
molecules are depleted from the gas phase due to low temperatures.
The dust emissivity is typically parameterized as a power-law function
of frequency, $\kappa_\nu = \kappa_0 (\nu/\nu_0)^\beta$.  The value of
$\beta$ is potentially a good measure of the size of the dust grains, with
$\beta = 2$ corresponding to grain sizes $a \ll \lambda$ where $\lambda$
is the observing wavelength.  This is typical for interstellar dust.
Lower values of $\beta$ are interpreted as evidence for larger dust
particles, and in the extreme case of $a \gg \lambda$ the dust opacity
is grey, with $\beta = 0$.  For distributions of particle sizes the
wavelength at which the slope of the dust opacity steepens indicates
the size of the largest particles present.

If the dust emission is optically thin and in the Rayleigh-Jeans (RJ)
part of the spectrum it is straightforward to derive $\beta$ from the
spectral index.  In this case, the flux density $F_\nu \propto \nu^2
(1-e^{-\tau}) \propto \nu^\alpha$, where $\alpha = 2+\beta$.  Corrections
for not being in the RJ are typically needed, and for the high column
densities expected in T Tauri disks the optically-thin approximation may
also not hold.  This is obviously a problem if $\alpha$ is to be used to
derive $\beta$, since high optical depth can mimic a low value of $\beta$.
Key to interpreting the spectral index of dust emission at millimeter
and submillimeter wavelengths is therefore the ability to resolve the
emission, in order to separate optical depth from $\beta$.  Furthermore,
in order to demonstrate particle growth to cm-sized planetesimals or
larger observations at the longest possible wavelength are needed.

\begin{figure}[!ht]
\plotfiddle{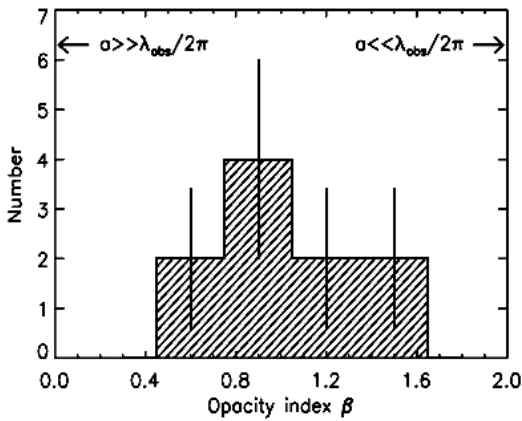}{2.1in}{0}{50}{50}{-150}{-120}
\caption{Distribution of dust opacity index, $\beta$, for 10 T Tauri stars
imaged at $\lambda = 7$~mm by \citet{chandler:Rodmann2006}.  The peak
of the distribution at $\beta \sim 1$ lies between that observed for
micron-sized dust grains in the ISM ($\beta \sim 2$) and the grey-body
opacity of particles with $a \gg \lambda$ ($\beta \sim 0$), indicating
the presence of cm-sized particles.}
\label{chandler:beta_dist}
\end{figure}

Several recent studies have been made using the VLA at 7~mm to investigate
the growth to cm-sized particles by resolving the emission from T Tauri
disks.  \citet{chandler:Rodmann2006} observed 14 low-mass, pre-main
sequence stars in the Taurus-Auriga star forming region, with a spatial
resolution of $1.5''$, corresponding to approximately 200~AU\@.  Ten were
detected at $5\sigma$ or better, and all were resolved in at least one
dimension.  After applying RJ corrections and for possible contributions
from free-free emission at this wavelength the distribution of $\beta$
(Figure~\ref{chandler:beta_dist}) indicates a peak at $\beta \sim 1$, and
shows evidence for cm-sized particles in these disks.  Similar results
have been found by \citet{chandler:Natta2004} for intermediate-mass,
pre-main sequence Herbig Ae stars, where 4 out of 9 disks observed
were resolved.

\begin{figure}[!ht]
\plotfiddle{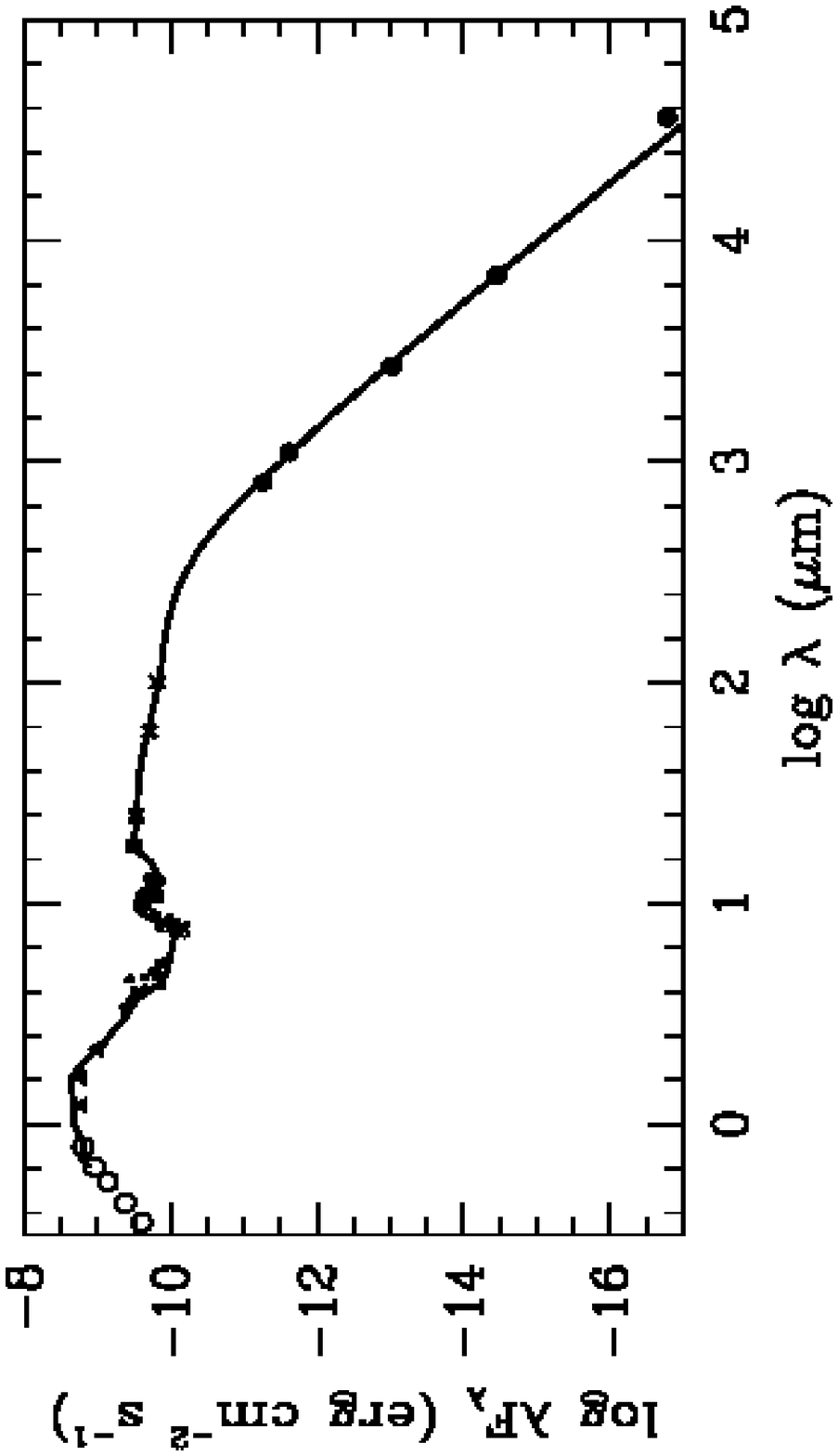}{2in}{270}{45}{45}{-175}{220}
\plotfiddle{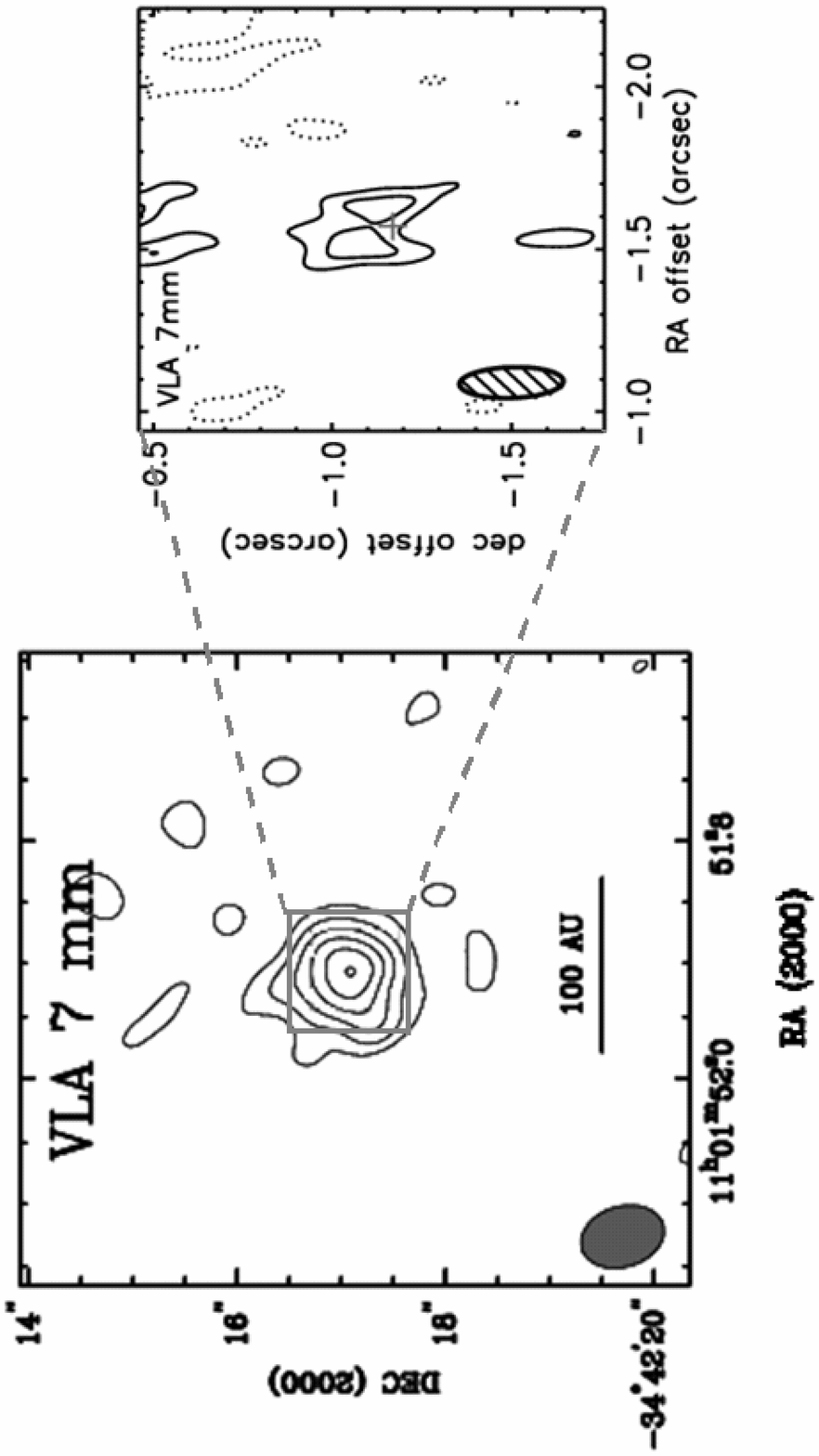}{2.4in}{270}{45}{45}{-175}{220}
\caption{{\it Top:} Spectral energy distribution of TW Hya shows a deficit
of emission at mid-infrared wavelengths indicative of a gap in the disk
\citep{chandler:Calvet2002}.  {\it Bottom, left:} 7~mm continuum emission
from TW Hya is resolved on scales of 10s of AU by the VLA in its most
compact, D, configuration \citep{chandler:Wilner2000}.  {\it Bottom,
right:} the central hole suggested by the SED is imaged by the VLA in
its most extended, A, configuration \citep{chandler:Hughes2007}.}
\label{chandler:tw_hya_fig}
\end{figure}

Recent observations of the 10~Myr old T Tauri star TW Hya provide evidence
of both planetesimal growth and disk clearing, the latter possibly due
to the presence of a planet.  The 7~mm continuum emission from this disk
was first resolved on size scales of 10s of AU by the VLA in its most
compact, D, configuration \citep[][Figure~\ref{chandler:tw_hya_fig},
bottom left]{chandler:Wilner2000}.  Subsequent modelling showed
that a dip in its spectral energy distribution at $\sim 10\mu$m,
and enhanced emission at $\sim 20\mu$m, was best accounted
for by an inner hole extending to 4~AU, and direct heating of
the inner edge of the disk at this radius by the central star
\citep[][Figure~\ref{chandler:tw_hya_fig}, top]{chandler:Calvet2002}.
This hole has now been imaged by the VLA in its most extended, A,
configuration \citep[][Figure~\ref{chandler:tw_hya_fig}, bottom
right]{chandler:Hughes2007}.

Evidence for cm-sized particles in the TW Hya disk comes
from 3.5~cm emission imaged with the VLA in its A configuration
\citep{chandler:Wilner2005}.  The emission is resolved with a brightness
temperature $\sim 10$~K, and is consistent with dust emission.
Further, the 3.5~cm emission is enhanced relative to that expected
for a ``standard,'' continuous power-law distribution of grain sizes,
but can be modelled as a bimodal grain size size distribution with some
fraction of interstellar-type dust and an extra component of ``pebbles.''
Considerable amounts of mass could be hidden in large particles in
disks in this way, the only evidence for which comes from observations
at cm-wavelengths.

\section{Disk kinematics from molecular line emission}

The primary tracer of circumstellar disk kinematics is emission from the
most abundant isotopomer of CO, $^{12}$CO\@.  Interferometric observations
of the emission in the $J$=1--0, 2--1, and 3--2 rotational transitions
at 115, 230, and 345~GHz respectively have been used to show that in
many cases the velocity structure in T Tauri disks is consistent with
Keplerian rotation \citep*{chandler:Simon2000, chandler:Dutrey2003,
chandler:Qi2004, chandler:Isella2007}.  Indeed, modelling of cubes of
spectral line emission can give the disk inclination, radial velocity
profile, and provide independent and direct determinations of the central
stellar mass to better than 10\% \citep[see, e.g.,][]{chandler:Simon2000}.
These methods are able to constrain theoretical pre-main sequence tracks,
and are now limited by the uncertain distance determinations for the
pre-main sequence stars.  However, parallax measurements from radio
VLBI observations are beginning to provide some of the most accurate
distances known for young stars exhibiting non-thermal emission, and
the distance to T Tau S has recently been pinned down to 0.4\% using
the VLBA \citep{chandler:Loinard2005, chandler:Loinard2007}.

\begin{figure}[!ht]
\plotfiddle{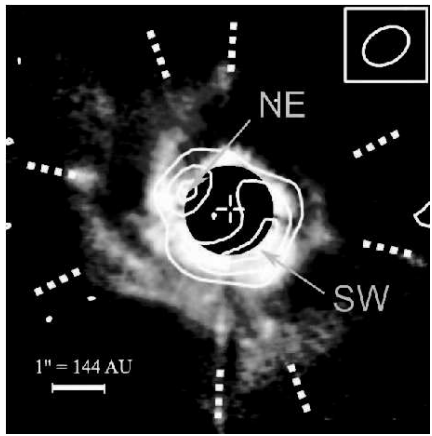}{2.3in}{0}{35}{35}{-100}{-50}
\caption{Small-scale dust emission at $\lambda = 0.87$~mm (contours)
overlaid on a coronographic near-infrared image (greyscale) of the young
star AB Aur from \citet{chandler:Lin2006}.  Both the near-infrared
and the dust emission illustrate spiral features, and the mm-wave
emission shows an inner hole in the disk at the position of the star.
The velocity structure of the CO emission from the disk is inconsistent
with Keplerian rotation.}
\label{chandler:ab_aur_fig}
\end{figure}

While most disks around pre-main sequence stars studied to date
do show Keplerian velocity profiles there is some evidence of
deviations from Keplerian rotation for the young star AB Aur
\citep*{chandler:Pietu2005,chandler:Lin2006}.  This source shows
molecular gas traced by $^{13}$CO(1--0) emission associated with a
spiral feature at optical wavelengths \citep*{chandler:Corder2005},
on scales of $\sim 4''$ ($\sim 600$~AU).  Both $^{12}$CO(3--2) and
dust continuum emission show a possible spiral feature on even smaller
scales \citep[][Figure~\ref{chandler:ab_aur_fig}]{chandler:Pietu2005,
chandler:Lin2006}.  \citet{chandler:Pietu2005} also show that if a
power-law radial velocity profile is assumed the best fit is $V \propto
r^{-0.41\pm0.01}$, significantly shallower than Keplerian.  As a result,
estimates of the stellar mass derived {\it assuming} Keplerian rotation
vary dramatically depending on the size scale and resolution of the
observations used.  Adding to the complexity, there is an inner hole of
radius $\sim 70$~AU \citep{chandler:Pietu2005}, and some indications of
outward radial motion along the spiral features \citep{chandler:Pietu2005,
chandler:Lin2006}.

Various possible explanations have been proposed for the observed
structure of AB Aur \citep{chandler:Pietu2005, chandler:Lin2006}.
Perhaps most attractive is the idea that there may be a low mass companion
in an inclined orbit that can produce both the inner hole and excite the
spiral features.  An encounter with a nearby star is also a possibility.
Alternatively, AB Aur remains surrounded by an extended circumstellar
envelope and also exhibits some evidence that its dust is not as evolved
as has been observed in other proto-planetary disks, suggesting that
the disk may just be sufficiently young that is has not yet relaxed to
a Keplerian configuration.

\section{Accretion and outflow}

\begin{figure}[!ht]
\plotfiddle{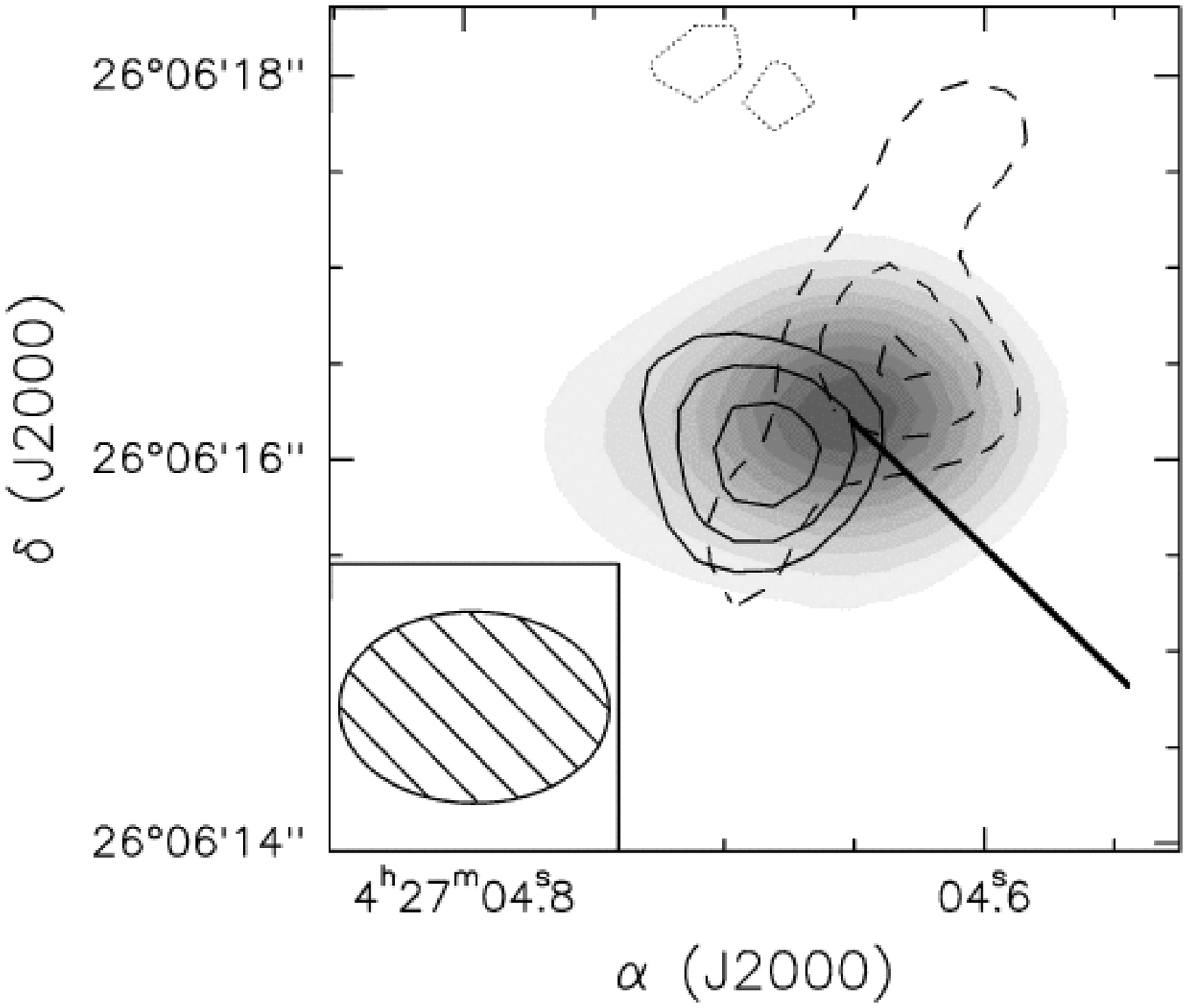}{1.9in}{0}{35}{35}{-205}{-80}
\plotfiddle{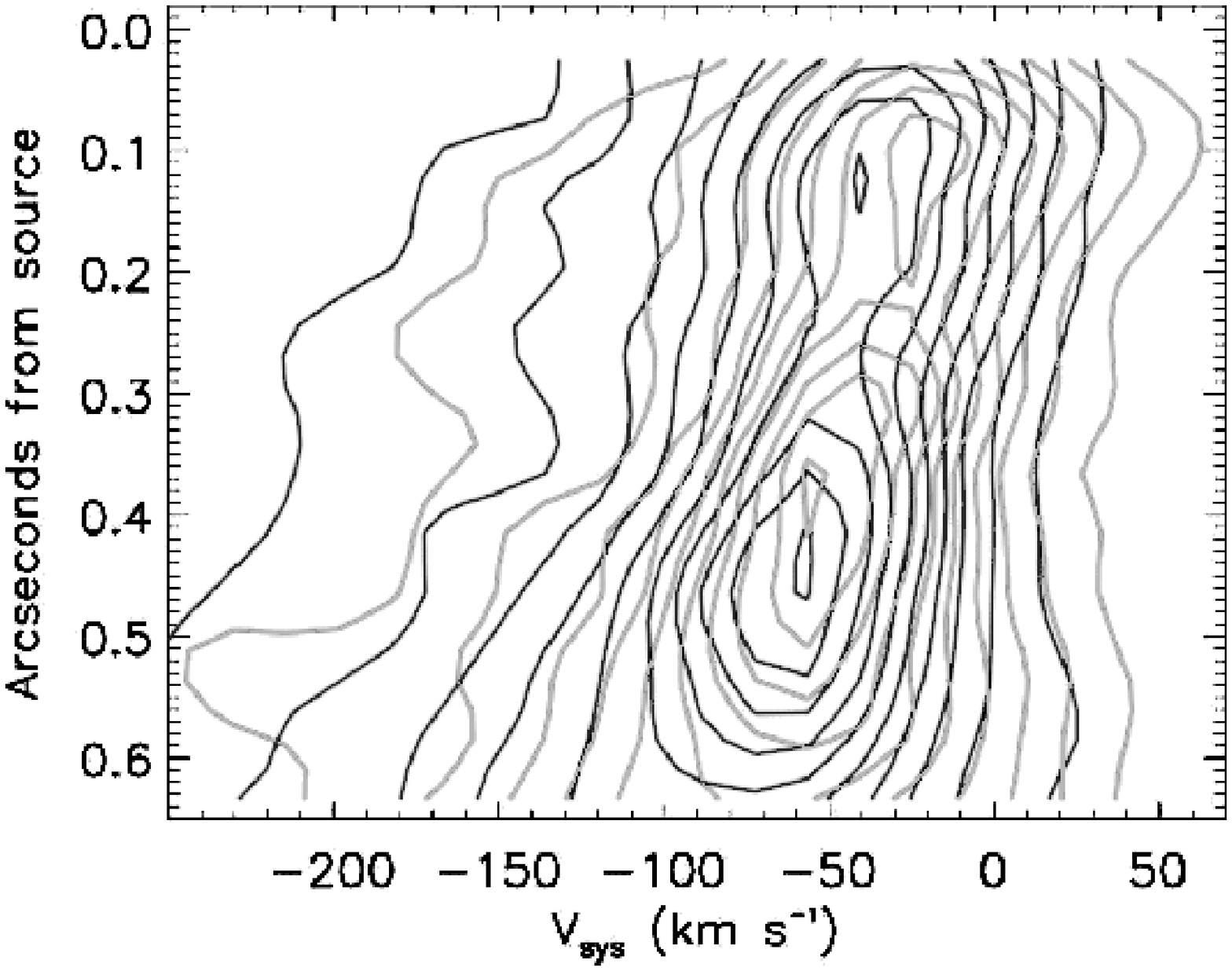}{0in}{0}{35}{35}{-15}{-55}
\caption{{\it Left:} Redshifted (dashed contours) and blueshifted (solid
contours) of the $^{13}$CO(2--1) emission from the DG Tau circumstellar
disk illustrating the direction of its rotation, overlaid on the 1.3~mm
dust continuum emission \citep{chandler:Testi2002}.  The thick solid
line gives the direction of the optical blueshifted jet.  {\it Right:}
[SII] emission in a slit offset 0.14$''$ to the southeast of the jet
axis (black contours) and 0.14$''$ to the northwest (grey contours),
from \citet{chandler:Bacciotti2002}.  There is a velocity offset between
the two slits in a direction consistent with the disk rotation.}
\label{chandler:dg_tau_fig}
\end{figure}

If accretion is to continue throughout the protostellar and T Tauri
phases of young, low-mass stars, at the rates implied by the observed
luminosities of those systems, some means of removing the angular
momentum from the accreting material is needed.  As is the case for
accretion disks in almost all astrophysical environments, jets and winds
are believed to be key to carrying away the angular momentum allowing
accretion to continue.  Measurements of rotation in protostellar
jets have been very difficult to make, however.  Perhaps the best
example to date is for DG Tauri (Figure~\ref{chandler:dg_tau_fig}).
The direction of rotation in its disk is traced by $^{13}$CO(2--1)
emission \citep{chandler:Testi2002}.  HST observations of the [SII] and
[OI] emission from the jet, in slits oriented at offsets $\pm0.14''$
from the jet axis, show a velocity gradient in the same sense as the
disk rotation \citep{chandler:Bacciotti2002}.  This may be the first
direct evidence of a disk/jet connection and the transport of angular
momentum using jets.

In spite of the sparse direct observational evidence, linked accretion
and outflow are fundamental ingredients in theories of star and planet
formation.  Outflow removes excess angular momentum from accreting gas,
the disk and wind regulate the accretion, and of course the remnant
disk ultimately becomes a planetary system.  Models for how this takes
place in detail fall into two broad categories: disk winds and X-winds.
Disk winds are winds originating at a range of radii in the disk, and most
models invoke rotating magnetic fields as the means by which material is
transferred from the disk to a collimated wind or jet moving perpendicular
to the disk surface \citep[see][for a review]{chandler:Konigl2000}.
In X-winds the wind originates at the ``X-point,'' the radius at which
magnetic field lines tied to the central star meet the inner radius of
the disk, which is therefore also the disk radius co-rotating with the
star \citep[e.g.,][]{chandler:Shu2000}.  Assuming the jet co-rotates
with a Keplerian disk, measured velocities imply the jet launching
point is spread over a range of radii from about 0.1~AU to up to a few AU
\citep[e.g.,][]{chandler:Coffey2007, chandler:Tatulli2007}.  These results
tend to favour disk winds over X-winds, but are not yet conclusive.

\section{Disks and massive star formation}

A reasonably well-established and understood picture of low-mass star
formation has emerged over the last two decades, including the vital
role of disks in mediating accretion and providing the building blocks
for planetary systems.  However, a simple scaling-up of this picture
to higher mass stars faces a problem for stars more massive than $\sim
10 M_\odot$.  Radiation pressure on infalling dust grains will reverse
the infall, unless the dust absorption cross-section per unit mass can
be significantly reduced (for example, by making the accreting material
very optically thick, perhaps by having a very high accretion rate),
or by reducing the effective luminosity by making the radiation field
anisotropic \citep[e.g.,][]{chandler:Yorke2002}.  Thus the properties of
massive protostars forming via disk-mediated accretion may be expected
to be somewhat different from their low-mass counterparts.

Direct observations of solar system-sized disks ($\sim 100$~AU) toward
early B stars are not as well established as those around T Tauri stars.
\citet{chandler:Shepherd2001} inferred the existence of a 130~AU disk
around a B2 young stellar object based on 7~mm continuum emission.
However, the disk was not resolved, and a model was needed to isolate the
disk emission (presumably dust) from ionized gas emission in the outflow
and surrounding HII region.  Further, evidence of rotation from molecular
line emission was not available on these size scales.  Several examples
of 1000 to 2000~AU flattened, rotating structures have been reported
around several early B (proto)stars which probably either trace the outer
accretion disk/dense torus that may be feeding the inner disk, or are the
result of confusion of multiple objects within the available resolution of
the observations: e.g., IRAS 20126+4104 \citep{chandler:Cesaroni1999,
chandler:Cesaroni2005}, Cep A HW2 \citep{chandler:Brogan2007,
chandler:Comito2007}, IRAS 23151+5912 \citep{chandler:Beuther2007}.
Given that these (proto)stars are still actively powering outflows, it is
likely that an inner, as yet undetectable, accretion disk is present that
launches the outflow.  Above a mass of about $15 M_\odot$, observations
of disks and disk winds/outflows are needed to verify whether they may
also form via disk-mediated accretion.

One of the complications with attempts at direct imaging of disks
around massive stars is that the nearest massive star-forming region,
the Orion Molecular Cloud at $D \sim 450$~pc, is a factor of ten more
distant than young, low-mass stars such as TW Hya, so that achieving the
necessary spatial resolution can be a challenge.  Masers provide a unique
opportunity in this respect, since their high non-thermal brightness
temperatures enable VLBI techniques to provide sub-milliarcsec resolution
in some cases.  Radio source I in Orion is associated with OH, H$_2$O,
and SiO masers, and the SiO masers in particular trace hot, dense gas
close to the protostar (the vibrational ground state traces densities
$n \sim 10^6$~cm$^{-3}$, temperatures $T \sim 1000$~K, while the first
vibrationally excited state traces $n \sim 10^{10\pm1}$~cm$^{-3}$, $T \sim
1000$--2000~K)\@.  Multi-epoch imaging of these masers using the VLBA have
recently measured the 3-dimensional dynamics of the disk/wind interaction
in the protostar \citep{chandler:GreenhillXX, chandler:MatthewsXX}.

\begin{figure}[!ht]
\plotfiddle{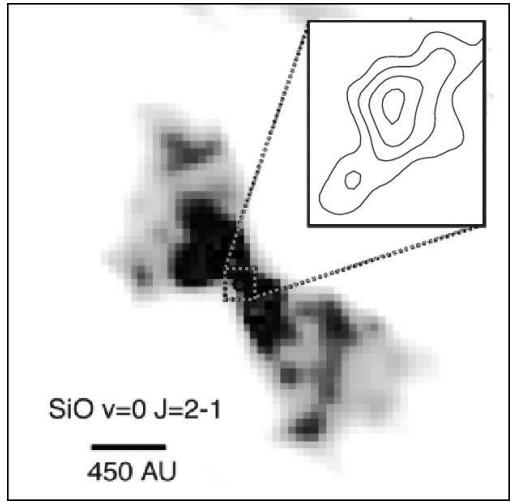}{2.6in}{0}{35}{35}{-100}{-40}
\caption{SiO $v$=0, $J$=2--1 emission from source I \citep[greyscale,
from][]{chandler:Wright1995} outlining a bowtie structure associated with
the outflow.  {\it Insert:} contours of 7~mm continuum emission trace
a disk oriented perpendicular to the outflow
\citep{chandler:Reid2007}.}
\label{chandler:bowtie_fig}
\end{figure}

\begin{figure}[!ht]
\plotfiddle{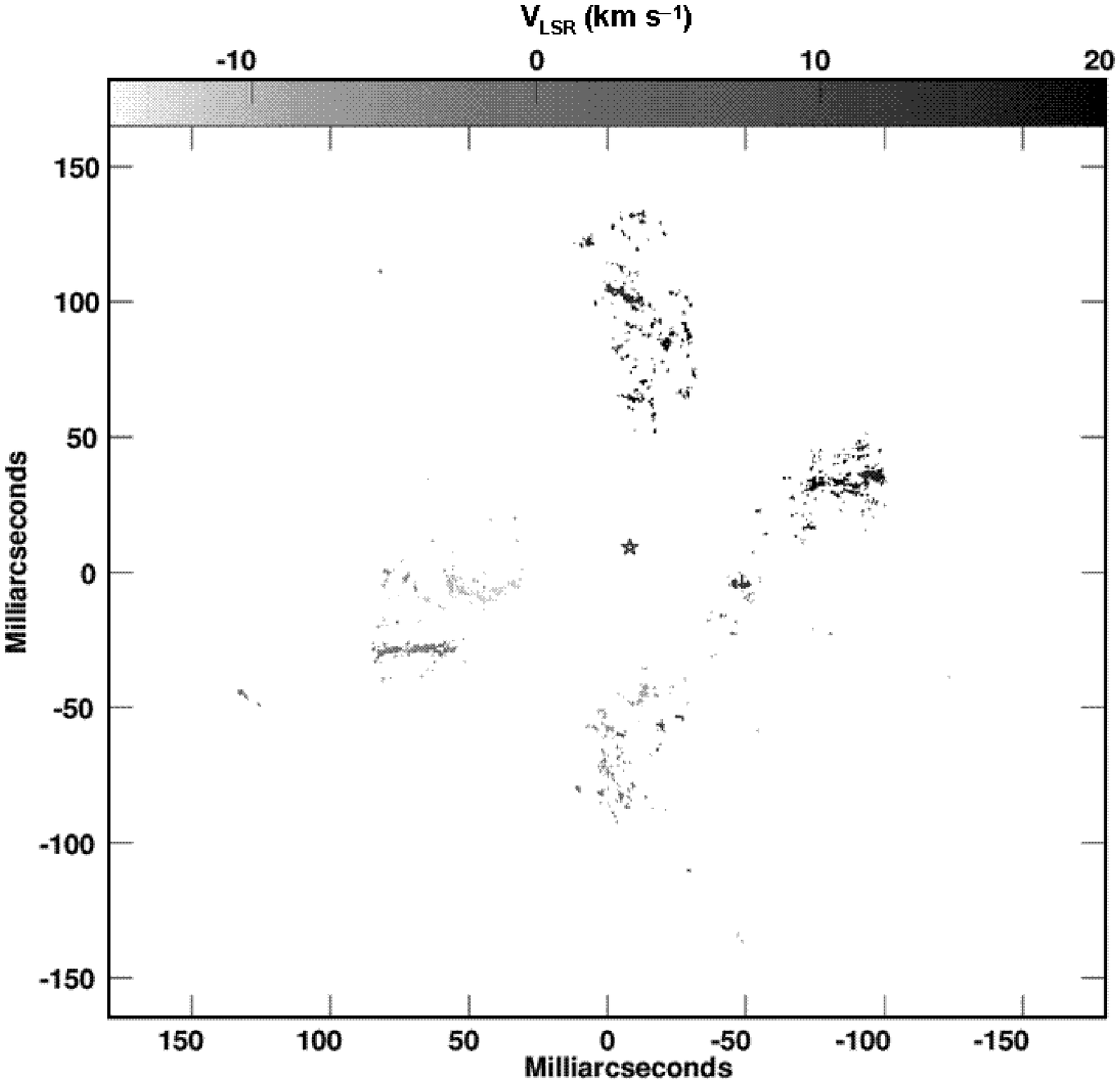}{1.9in}{0}{35}{35}{-210}{-80}
\plotfiddle{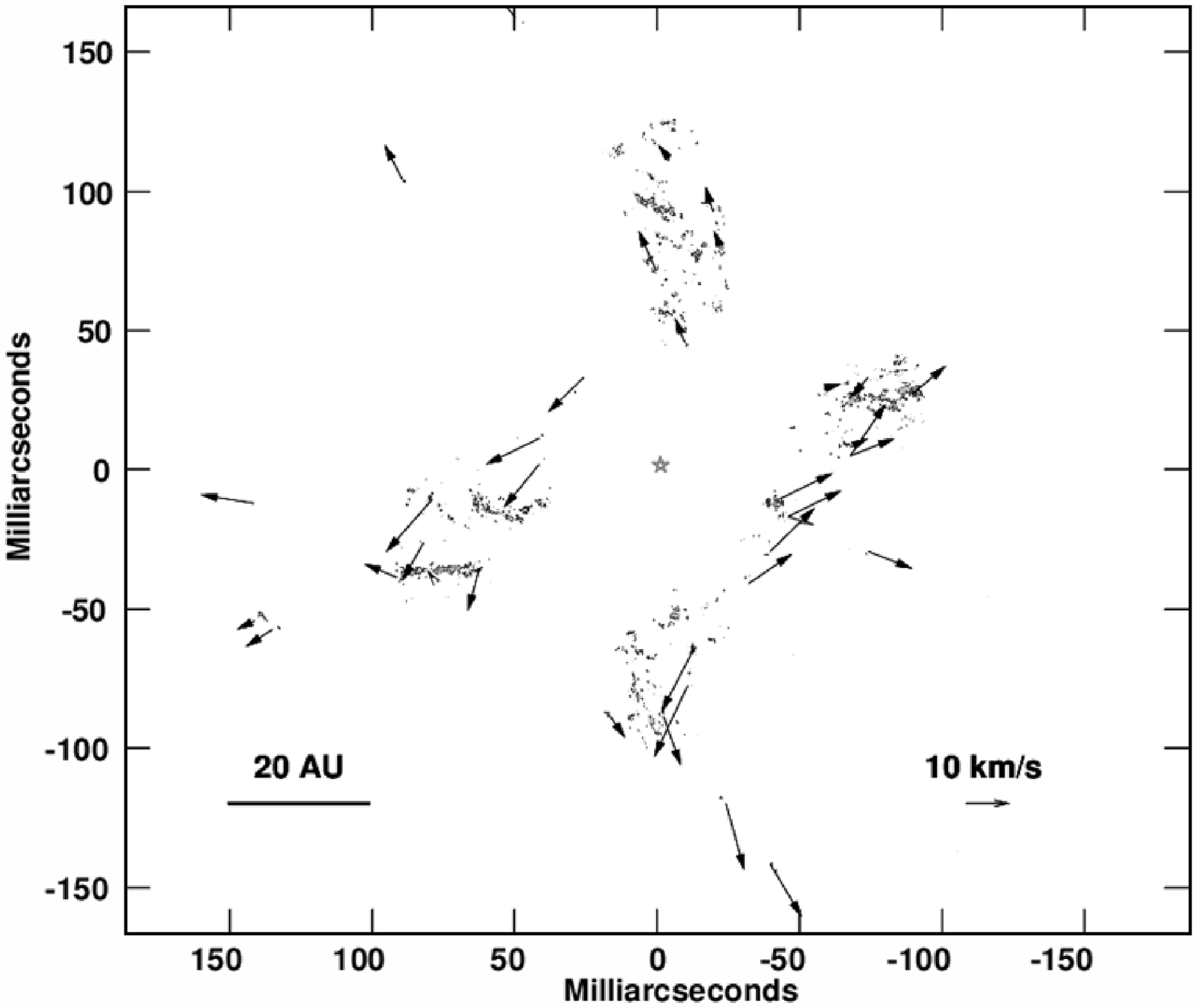}{0in}{0}{35.5}{35.5}{-20}{-66.5}
\caption{{\it Left:} Radial velocity derived from $v$=1 and
$v$=2, $J$=1--0 SiO maser emission for one observational epoch
\citep{chandler:GreenhillXX}.  The star represents the position of source
I determined from radio continuum emission.  {\it Right:} Proper motions
of the SiO masers derived from observations spanning multiple epochs
\citep{chandler:MatthewsXX}.  The combination of the radial velocities and
proper motions are best explained by a rotating, almost edge-on disk, with
the SiO masers tracing the interaction of a wind with the disk surface.}
\label{chandler:sourceI_masers_fig}
\end{figure}

The vibrational ground state SiO emission traces a bow tie
structure, while the 7~mm continuum emission appears to trace a disk
(Figure~\ref{chandler:bowtie_fig}).  The vibrationally excited masers
predominantly outline a cross centered on the continuum, but there are
also masers bridging the gap between the southern and western arms of the
cross.  The best model consistent with the remarkably systematic velocity
patterns in the maser emission (Figure~\ref{chandler:sourceI_masers_fig},
left) is that of an almost edge-on disk, with the SiO masers tracing
the interaction of a wind with the disk surface, and subsequent
collimation of the wind by magnetic fields and the surrounding
molecular cloud.  Proper motions show that both SiO and H$_2$O
masers are moving approximately perpendicular to the disk surface
(Figure~\ref{chandler:sourceI_masers_fig}, right).  The diameter of the
disk is $\sim 150$~mas ($\sim 70$~AU), and the resolution with which
the masers have been observed is $\sim 0.4$~mas ($\sim 0.2$~AU).

\section{The future of disk observations: ALMA and EVLA}

The resolution and sensitivity of two new instruments that will become
available within the next five years, the Atacama Large Millimeter
Array (ALMA) and the Expanded Very Large Array (EVLA), will completely
revolutionize the study of circumstellar disks.  ALMA will operate at
wavelengths between 9~mm and 320~$\mu$m, with up to 64 12-m antennas and
resolutions as small as 0.01$''$ at the shortest wavelengths.  It will be
superb for resolving disk structure and kinematics for T Tauri stars and
debris disks.  Its 8~GHz of bandwidth and excellent continuum sensitivity
makes it ideal for studying disk chemistry and the direct detection
of proto-planets.

The EVLA will have 8~GHz of bandwidth at 1.3~cm and 7~mm, providing a
factor of ten improvement in continuum sensitivity over the VLA and a
factor of 2 to 4 better angular resolution than ALMA will be able to
achieve at the equivalent wavelength.  Long wavelength observations
are vital for demonstrating planetesimal formation, and for imaging
disk formation in the youngest protostars where the dust emission will
be optically thick at the wavelengths where ALMA will have the highest
spatial resolution.

\section{Conclusions}

Over the last few years, submm, mm, and cm-wave observations of
circumstellar disks around low-mass, pre-main sequence stars have
demonstrated evidence for grain growth to cm-sized particles, a
necessary precursor to planet formation.  Keplerian velocity fields in
those disks have enabled independent measurements of central stellar
masses, and are able to constrain pre-main sequence evolutionary tracks.
Deviations from Keplerian rotation have been observed in some cases, and
raise the possibility that low-mass companions may be inferred from such
observations.  Evidence for angular momentum transport in disk winds has
been demonstrated by the first observations of rotation in stellar jets,
in a sense consistent with that of the accompanying disk.  Masers are
now able to trace disk and wind interactions for a massive protostar,
suggesting that disk-mediated accretion may be a mechanism for forming
high-mass stars.  The future of disk studies is very bright; ALMA and
EVLA will transform this field by providing images of gaps in disks,
detecting dust heated by proto-planets, revealing the early phases of
disk formation, and through detailed studies of the gas chemistry.

\acknowledgements 

The National Radio Astronomy Observatory is a facility of the National
Science Foundation operated under cooperative agreement by Associated
Universities, Inc.


\begin{thebibliography}{}

\bibitem[Bacciotti et al. (2002)]{chandler:Bacciotti2002}Bacciotti, F.,
Ray, T. P., Mundt, R., Eisl\"offel, J., \& Solf, J. 2002, ApJ, 576, 222

\bibitem[Beckwith et al. (1990)]{chandler:Beckwith1990}Beckwith, S. V. W.,
Sargent, A. I., Chini, R. S., \& G\"usten, R. 1990, AJ, 99, 924

\bibitem[Beuther et al. (2007)]{chandler:Beuther2007}Beuther, H., Zhang,
Q., Hunter, T. R., Sridharan, T. K., \& Bergin, E. A. 2007, A\&A, 473, 493

\bibitem[Brogan et al. (2007)]{chandler:Brogan2007}Brogan, C. L.,
Chandler, C. J., Hunter, T. R., Shirley, Y. L., \& Sarma, A. P. 2007,
ApJ, 660, L133

\bibitem[Calvet et al. (2002)]{chandler:Calvet2002}Calvet, N., D'Alessio,
P., Hartmann, L., Wilner, D., Walsh, A., \& Sitko, M. 2002, ApJ, 568, 1008

\bibitem[Cesaroni et al. (1999)]{chandler:Cesaroni1999}Cesaroni, R.,
Felli, M., Jenness, T., Neri, R., Olmi, L., Robberto, M., Testi, L., \&
Walmsley, C. M. 1999, A\&A, 345, 949

\bibitem[Cesaroni et al. (2005)]{chandler:Cesaroni2005}Cesaroni, R.,
Neri, R., Olmi, L., Testi, L., Walmsley, C. M., \& Hofner, P. 2005,
A\&A, 434, 1039

\bibitem[Coffey et al. (2007)]{chandler:Coffey2007}Coffey, D., Bacciotti,
F., Ray, T. P., Eisl\"offel, J., \& Woitas, J. 2007, ApJ, 663, 350

\bibitem[Comito et al. (2007)]{chandler:Comito2007}Comito, C., Schilke,
P., Endesfelder, U., Jim\'enez-Serra, I., \& Mart\'\i n-Pintado, J. 2007,
A\&A, 469, 207

\bibitem[Corder, Eisner, \& Sargent (2005)]{chandler:Corder2005}Corder,
S., Eisner, J., \& Sargent, A. 2005, ApJ, 622, L133

\bibitem[Dutrey et al. (2003)Dutrey, Guilloteau, \&
Simon]{chandler:Dutrey2003}Dutrey, A., Guilloteau, S., \& Simon, M. 2003,
A\&A, 402, 1003

\bibitem[Greenhill et al., in prep.]{chandler:GreenhillXX}Greenhill,
L. J., et al., in preparation

\bibitem[Hughes et al. (2007)]{chandler:Hughes2007}Hughes, A. M., Wilner,
D.  J., Calvet, N., D'Alessio, P., Claussen, M. J., \& Hogerheijde, M.
R. 2007, ApJ, 664, 536

\bibitem[Isella et al. (2007)]{chandler:Isella2007}Isella, A., Testi,
L., Natta, A., Neri, R., Wilner, D., \& Qi, C. 2007, A\&A, 469, 213

\bibitem[K\"onigl \& Pudritz (2000)]{chandler:Konigl2000}K\"onigl, A.,
\& Pudritz, R. E. 2000, in Protostars and Planets IV, ed.\ V. Mannings,
A. P. Boss, \& S. S. Russell (Tucson: University of Arizona Press), 759

\bibitem[Lin et al. (2006)]{chandler:Lin2006}Lin, S.-Y., Ohashi, N.,
Lim, J., Ho, P. T. P., Fukagawa, M., \& Tamura, M. 2006, ApJ, 645, 1297

\bibitem[Loinard et al. (2005)]{chandler:Loinard2005}Loinard, L.,
Mioduszewski, A. J., Rodr\'\i guez, L. F., Gonz\'alez, R. A., Rod\'\i
guez, M. I., \& Torres, R. M. 2005, ApJ, 619, L179

\bibitem[Loinard et al. (2007)]{chandler:Loinard2007}Loinard, L., et
al. 2007, ApJ, 671, 546

\bibitem[Matthews et al., in prep.]{chandler:MatthewsXX}Matthews, L.
D., et al., in preparation

\bibitem[Natta et al. (2004)]{chandler:Natta2004}Natta, A., Testi, L.,
Neri, R., Shepherd, D. S., \& Wilner, D. J. 2004, A\&A, 416, 179

\bibitem[Pi\'etu et al. (2005)Pi\'etu, Guilloteau, \&
Dutrey]{chandler:Pietu2005}Pi\'etu, V., Guilloteau, S., \& Dutrey,
A. 2005, A\&A, 443, 945

\bibitem[Qi et al. (2004)]{chandler:Qi2004}Qi, C., et al.\ 2004, ApJ,
616, L11

\bibitem[Reid et al. (2007)]{chandler:Reid2007}Reid, M. J., Menten,
K. M., Greenhill, L. J., \& Chandler, C. J. 2007, ApJ, 664, 950

\bibitem[Rodmann et al. (2006)]{chandler:Rodmann2006}Rodmann, J., Henning,
T., Chandler, C. J., Mundy, L. G., \& Wilner, D. J. 2006, A\&A, 446, 211

\bibitem[Shepherd et al. (2001)Shepherd, Claussen, \&
Kurtz]{chandler:Shepherd2001}Shepherd, D. S., Claussen, M. J., \& Kurtz,
S. E. 2001, Science, 292, 1513

\bibitem[Shu et al. (2000)]{chandler:Shu2000}Shu, F. H., Najita, J. R.,
Shang, H., \& Li, Z.-Y. 2000, in Protostars and Planets IV, ed.\ V.
Mannings, A. P. Boss, \& S. S. Russell (Tucson: University of Arizona
Press), 789

\bibitem[Simon et al. (2000)Simon, Dutrey, \&
Guilloteau]{chandler:Simon2000}Simon, M., Dutrey, A., \& Guilloteau,
S. 2000, ApJ, 545, 1034

\bibitem[Tatulli et al. (2007)]{chandler:Tatulli2007}Tatulli, E., et al.,
2007, A\&A, 464, 55

\bibitem[Testi et al. (2002)]{chandler:Testi2002}Testi, L., Bacciotti,
F., Sargent, A. I., Ray, T. P., \& Eisl\"offel, J. 2002, A\&A, 394, L31

\bibitem[Wilner et al. (2000)]{chandler:Wilner2000}Wilner, D. J., Ho,
P. T. P., Kastner, J. H., \& Rodr\'\i guez, L. F. 2000, ApJ, 534, L101

\bibitem[Wilner et al. (2005)]{chandler:Wilner2005}Wilner, D. J.,
D'Alessio, P., Calvet, N., Claussen, M. J., \& Hartmann, L. 2005, ApJ,
626, L109

\bibitem[Wright et al. (1995)]{chandler:Wright1995}Wright, M. C. H.,
Plambeck, R. L., Mundy, L. G., \& Looney, L. W. 1995, ApJ, 455, L185

\bibitem[Yorke (2002)]{chandler:Yorke2002}Yorke, H. W. 2002, in Hot
Star Workshop III: The Earliest Stages of Massive Star Birth, ASP Conf.\
Ser.\ 267, ed.\ P. A. Crowther (San Francisco: ASP), 165

\end{thebibliography}
\end{document}